\documentclass[12pt]{article}
\usepackage{amsfonts}
\usepackage{amsmath}
\usepackage{amssymb}

\input{tcilatex}
\begin{document}

\begin{center}
\smallskip \ 

\textbf{HIGHER DIMENSIONAL ELKO THEORY}

\smallskip \ 

\smallskip \ 

J. A. Nieto \footnote{%
nieto@uas.edu.mx, janieto1@asu.edu}

\smallskip \ 

\smallskip \ 

\smallskip \ 

\textit{Facultad de Ciencias F\'{\i}sico-Matem\'{a}ticas de la Universidad
Aut\'{o}noma} \textit{de Sinaloa, 80010, Culiac\'{a}n Sinaloa, M\'{e}xico.}

\smallskip \ 

\smallskip \ 

\textbf{Abstract}
\end{center}

We show that the so called Elko equation can be derived from a $5$%
-dimensional Dirac equation. We argue that this result can be relevant for
dark matter and cosmological scenarios. We generalize our procedure to
higher dimensions.

\bigskip \ 

\bigskip \ 

\bigskip \ 

\bigskip \ 

\bigskip \ 

\bigskip \ 

\bigskip \ 

\bigskip \ 

\bigskip \ 


Pacs numbers: 04.20.Gz, 04.60.-Ds, 11.30.Ly

July, 2014

\newpage

\noindent It is known that one of the most interesting proposals to explain
dark matter [1] is provided by the so called Elko theory [2-3] (see also
Refs. [4]-[12]). This theory describes spin half-integer fermions with dual
helicity eigenspinors of the charge conjugation operator. It turns out that
the growing interest in this kind of matter is due to the fact that besides
being a candidate for dark matter it may also provide an alternative
explanation of the accelerated expansion our universe [13]-[15] (see also
Refs. [2], [16], and [17] references therein). In the context of
inequivalent spin structures on arbitrary curved spacetimes exotic dark
spinor fields has been introduced [18]. Moreover, dynamical dispersion
relations for Elko dark spinors fields have lead to mass generation proposal
[19] and a light Elko signals exploration at acelerators has been developed
[20]. It is interesting that Elko spinor fields have been considered as a
tool for probing exotic topological spacetime features [21]. Even more
recently, a Lagrangian for mass dimension one fermion has been derived [22].

Recently, it has been proposed [23] a $5$-dimensional Elko theory in the
context of Minkowski branes and it was shown that if a $5$-dimensional mass
Kaluza-Klein term (see Ref. [24] for details of Kaluza-Klein theory) is
introduced there is not the possibility to localize these modes on the
corresponding branes. Part of the motivation in these developments arises
from the desire to shed some light on the brane world theory that originates
from so-called $M$-theory (see Ref. [25] and references therein), which is a
generalization of superstring theory [26]. In this case, the brane is
embedded in a higher dimensional space-time. So it appears interesting to
explore whether the Elko theory is some how related to $M$-theory. As a
first step in this direction, one would like to explore whether there exists
a higher dimensional Elko theory. In this work we show the surprising result
that the Elko equation in four dimensions can be obtained from a higher
dimensional Dirac equation.

Let us start mentioning that the Elko theory is based on the fundamental
equation

\begin{equation}
\left[ \gamma ^{\mu }\hat{p}_{\mu }\delta _{a}^{b}+im_{0}\varepsilon _{a}^{b}%
\right] \psi _{b}=0.  \tag{1}
\end{equation}%
Here, the indices $\mu ,\nu $ etc. run from $0$ to $3$ and the indices $a,b$
run from $1$ to $2$. Further the $\gamma $-matrices satisfy the Clifford
algebra

\begin{equation}
\gamma ^{\mu }\gamma ^{\nu }+\gamma ^{\nu }\gamma ^{\mu }=-2\eta ^{\mu \nu },
\tag{2}
\end{equation}%
with $\eta ^{\mu \nu }=diag(-1,1,1,1)$. Moreover, $\hat{p}_{\mu }=-i\partial
_{\mu }$ and $\varepsilon ^{ab}$ is the completely antisymmetric $%
\varepsilon $-symbol, with $\varepsilon ^{12}=1=-\varepsilon ^{21}$.

In contrast to the usual Dirac equation [27]

\begin{equation}
\left[ \gamma ^{\mu }\hat{p}_{\mu }+m_{0}\right] \psi =0,  \tag{3}
\end{equation}%
equation (1) requires eight-component complex spinor $\psi _{b}$ rather than
four-component $\psi $ which is the case in equation (3). Furthermore, the
quantities $\delta _{a}^{b}$ and $\varepsilon _{a}^{b}$ in (1) establish
that $\psi _{b}$ is not eigenspinor of the $\gamma ^{\mu }\hat{p}_{\mu }$
operator as $\psi $ in equation (3). In spite of these key differences, one
can prove that (1) and (3) imply the Klein-Gordon equations

\begin{equation}
\left[ \hat{p}^{\mu }\hat{p}_{\mu }+m_{0}^{2}\right] \psi _{a}=0  \tag{4}
\end{equation}%
and%
\begin{equation}
\left[ \hat{p}^{\mu }\hat{p}_{\mu }+m_{0}^{2}\right] \psi =0,  \tag{5}
\end{equation}%
respectively. Therefore, both (1) and (3) can be understood as
\textquotedblleft the square root of the Klein-Gordon
equation.\textquotedblright

Now, consider a $5$-dimensional Klein-Gordon equation

\begin{equation}
\left[ \hat{p}^{\hat{\mu}}\hat{p}_{\hat{\mu}}+m_{0}^{2}\right] \psi =0, 
\tag{6}
\end{equation}%
or

\begin{equation}
\left[ (\hat{p}_{5})^{2}+\hat{p}^{\mu }\hat{p}_{\mu }+m_{0}^{2}\right] \psi
=0,  \tag{7}
\end{equation}%
with the indices $\hat{\mu}$, $\hat{\nu}$ etc. assume the values $0,1,2,3,5$%
. Here, $\hat{p}^{\hat{\mu}}\hat{p}_{\hat{\mu}}=\eta ^{\hat{\mu}\hat{\nu}}%
\hat{p}_{\hat{\mu}}\hat{p}_{\hat{\nu}}$, with $\eta ^{\hat{\mu}\hat{\nu}%
}=diag(-1,1,1,1,1)$.

By virtue of (2) one finds that (7) can be written as

\begin{equation}
\left[ -(-\hat{p}_{5}+\gamma ^{\mu }\hat{p}_{\mu })(\hat{p}_{5}+\gamma ^{\nu
}\hat{p}_{\nu })+m_{0}^{2}\right] \psi =0.  \tag{8}
\end{equation}%
Thus, introducing the definitions

\begin{equation}
\psi _{L}\equiv \psi  \tag{9}
\end{equation}%
and

\begin{equation}
\psi _{R}\equiv -\frac{i}{m_{0}}(\hat{p}_{5}+\gamma ^{\mu }\hat{p}_{\mu
})\psi _{L},  \tag{10}
\end{equation}%
one sees that (8) leads to

\begin{equation}
(-\hat{p}_{5}+\gamma ^{\mu }\hat{p}_{\mu })\psi _{R}+im_{0}\psi _{L}=0, 
\tag{11}
\end{equation}%
while, the expressions (10) can be rewritten as

\begin{equation}
(\hat{p}_{5}+\gamma ^{\mu }\hat{p}_{\mu })\psi _{L}-im_{0}\psi _{R}=0. 
\tag{12}
\end{equation}

Let us now assume that $\hat{p}_{5}\psi _{R}=0$ and $\hat{p}_{5}\psi _{L}=0$%
. (These imitate the cylindrical conditions in the usual dimensional
reduction of the Kaluza-Klein theory [24].) With these assumptions,
equations (11) and (12) become

\begin{equation}
\gamma ^{\mu }\hat{p}_{\mu }\psi _{R}+im_{0}\psi _{L}=0  \tag{13}
\end{equation}%
and

\begin{equation}
\gamma ^{\mu }\hat{p}_{\mu }\psi _{L}-im_{0}\psi _{R}=0,  \tag{14}
\end{equation}%
respectively. So, if one assumes that the indices $a,b$ run from the labels $%
R$ and $L$ one learns that (13) and (14) can be obtained from the expression

\begin{equation}
(\gamma ^{\mu }\hat{p}_{\mu }\delta _{a}^{b}+i\varepsilon _{a}^{b}m_{0})\psi
_{b}=0,  \tag{15}
\end{equation}%
which is precisely formula (1) (see formula (5.14) in Refs. [2]). This
means, among other things, that (11) and (12) represent a generalization of
Elko equation (1).

Now, it is straightforward to generalize our procedure to higher dimensions.
In fact, let us introduce the $D$-dimensional Klein-Gordon equation

\begin{equation}
\left[ \hat{p}^{A}\hat{p}_{A}+m_{0}^{2}\right] \psi =0,  \tag{16}
\end{equation}%
with the indices $A$, $B$ etc. running from $0$ to $D-1$. We shall assume
that the splitting of the indices $A=(\mu ,a)$. Thus, considering that $\hat{%
p}^{A}\hat{p}_{A}=\eta ^{AB}\hat{p}_{A}\hat{p}_{B}$, with $\eta
^{AB}=diag(-1,1,...,1,1)$ one sees that (16) become

\begin{equation}
\left[ \hat{p}^{a}\hat{p}_{a}+\hat{p}^{\mu }\hat{p}_{\mu }+m_{0}^{2}\right]
\psi =0.  \tag{17}
\end{equation}%
Using the analogue of (2) for the internal space%
\begin{equation}
\gamma ^{a}\gamma ^{b}+\gamma ^{b}\gamma ^{a}=2\delta ^{ab},  \tag{18}
\end{equation}%
and also assuming that

\begin{equation}
\gamma ^{a}\gamma ^{\mu }-\gamma ^{\mu }\gamma ^{a}=0,  \tag{19}
\end{equation}%
one learns that (17) can also be written as

\begin{equation}
\left[ -(-\gamma ^{a}\hat{p}_{a}+\gamma ^{\mu }\hat{p}_{\mu })(\gamma ^{b}%
\hat{p}_{b}+\gamma ^{\nu }\hat{p}_{\nu })+m_{0}^{2}\right] \psi =0.  \tag{20}
\end{equation}%
It is not difficult to see that the $\gamma ^{\mu }$ and $\gamma ^{a}$
matrices can be chosen as $2^{(D-2)/2}\times 2^{(D-2)/2}$ or $%
2^{(D-3)/2}\times 2^{(D-3)/2}$ matrices, depending if $D$ is an even or odd
number, respectively.

Now, let us introduce the definitions

\begin{equation}
\psi _{L}\equiv \psi  \tag{21}
\end{equation}%
and

\begin{equation}
\psi _{R}=-\frac{i}{m_{0}}(\gamma ^{a}\hat{p}_{a}+\gamma ^{\mu }\hat{p}_{\mu
})\psi _{L}.  \tag{22}
\end{equation}%
This definitions allow us to rewrite (20) as

\begin{equation}
-i(-\gamma ^{a}\hat{p}_{a}+\gamma ^{\mu }\hat{p}_{\mu })\psi _{R}+m_{0}\psi
_{L}=0.  \tag{23}
\end{equation}%
The Eqs. (22) and (23) lead to

\begin{equation}
(-\gamma ^{a}\hat{p}_{a}+\gamma ^{\mu }\hat{p}_{\mu })\psi _{R}+im_{0}\psi
_{L}=0  \tag{24}
\end{equation}%
and

\begin{equation}
(\gamma ^{a}\hat{p}_{a}+\gamma ^{\mu }\hat{p}_{\mu })\psi _{L}-im_{0}\psi
_{R}=0.  \tag{25}
\end{equation}%
So, once again, if one assumes the dimensional reductions conditions

\begin{equation}
\gamma ^{a}\hat{p}_{a}\psi _{R}=0  \tag{26}
\end{equation}%
and

\begin{equation}
\gamma ^{a}\hat{p}_{a}\psi _{L}=0,  \tag{27}
\end{equation}%
one obtains Elko formula (1). Of course, our approach of this higher
dimensional generalization of Elko equation resembles the typical procedure
of the Dirac equation in the Weyl representation. In this case one starts
with the Klein-Gordon equation%
\begin{equation}
(\eta ^{\mu \nu }\hat{p}_{\mu }\hat{p}_{\nu }+m_{0}^{2})\psi =0,  \tag{28}
\end{equation}%
and introduces the Pauli matrices,

\begin{equation}
\sigma ^{1}=\left( 
\begin{array}{cc}
0 & 1 \\ 
1 & 0%
\end{array}%
\right) \ ,\  \sigma ^{2}=\left( 
\begin{array}{cc}
0 & -i \\ 
i & 0%
\end{array}%
\right) \ ,\  \sigma ^{3}=\left( 
\begin{array}{cc}
1 & 0 \\ 
0 & -1%
\end{array}%
\right) ,  \tag{29}
\end{equation}%
which satisfy

\begin{equation}
\sigma ^{i}\sigma ^{j}+\sigma ^{j}\sigma ^{i}=2\delta ^{ij}  \tag{30}
\end{equation}%
and

\begin{equation}
\sigma ^{i}\sigma ^{j}-\sigma ^{j}\sigma ^{i}=2i\varepsilon ^{ijk}\sigma
_{k}.  \tag{31}
\end{equation}%
Here, $\delta ^{ij}=diag(1,1,1)$ is the Kronecker delta and $\varepsilon
^{ijk}$ is the completely antisymmetric $\varepsilon $-symbol, with $%
\varepsilon ^{123}=1$. In $(1+3)$-dimensions we have $\eta ^{\mu \nu
}=diag(-1,1,1,1)$ and therefore (28) becomes%
\begin{equation}
(-\hat{p}_{0}\hat{p}_{0}+\delta ^{ij}\hat{p}_{i}\hat{p}_{j}+m_{0}^{2})\psi
=0.  \tag{32}
\end{equation}%
Using (30) one sees that (32) can be written as

\begin{equation}
(-\sigma ^{0}\hat{p}_{0}\sigma ^{0}\hat{p}_{0}+\sigma ^{i}\sigma ^{j}\hat{p}%
_{i}\hat{p}_{j}+m_{0}^{2})\psi =0  \tag{33}
\end{equation}%
or

\begin{equation}
(-\sigma ^{0}\hat{p}_{0}+\sigma ^{i}\hat{p}_{i})(\sigma ^{0}\hat{p}%
_{0}+\sigma ^{j}\hat{p}_{j})\psi +m_{0}^{2}\psi =0,  \tag{34}
\end{equation}%
where $\sigma ^{0}$ is the identity $2\times 2$-matrix.

So, by defining $\psi _{L}\equiv \psi $ and

\begin{equation}
\psi _{R}\equiv -\frac{1}{m_{0}}(\sigma ^{0}\hat{p}_{0}+\sigma ^{j}\hat{p}%
_{j})\psi _{L},  \tag{35}
\end{equation}%
one obtains%
\begin{equation}
(\sigma ^{0}\hat{p}_{0}+\sigma ^{j}\hat{p}_{j})\psi _{L}+m_{0}\psi _{R}=0 
\tag{36}
\end{equation}%
and%
\begin{equation}
(\sigma ^{0}\hat{p}_{0}-\sigma ^{i}\hat{p}_{i})\psi _{R}+m_{0}\psi _{L}=0. 
\tag{37}
\end{equation}%
We recognize in (36) and (37) the Dirac equation in the Weyl representation,

\begin{equation}
\left[ \gamma _{W}^{0}\hat{p}_{0}+\gamma _{W}^{i}\hat{p}_{i}+m_{0}\right]
\psi _{W}=0,  \tag{38}
\end{equation}%
which in covariant notation becomes

\begin{equation}
\left[ \gamma _{W}^{\mu }\hat{p}_{\mu }+m_{0}\right] \psi _{W}=0,  \tag{39}
\end{equation}%
where%
\begin{equation}
\gamma _{W}^{0}=\left( 
\begin{array}{cc}
0 & 1 \\ 
1 & 0%
\end{array}%
\right) \  \ ,\  \  \gamma _{W}^{i}=\left( 
\begin{array}{cc}
0 & \sigma ^{i} \\ 
-\sigma ^{i} & 0%
\end{array}%
\right) ,  \tag{40}
\end{equation}%
and

\begin{equation}
\psi _{W}=\left( 
\begin{array}{c}
\psi _{R} \\ 
\psi _{L}%
\end{array}%
\right) .  \tag{41}
\end{equation}%
Here, the $W$ in $\gamma _{W}^{\mu }$ and $\psi _{W}$ means that this
quantities are in the Weyl representation. One can verify that 
\begin{equation}
\gamma _{W}^{\mu }\gamma _{W}^{\nu }+\gamma _{W}^{\nu }\gamma _{W}^{\mu
}=-2\eta ^{\mu \nu }.  \tag{42}
\end{equation}

It turns out that one can interchange the order of (36) and (37) in the form

\begin{equation}
(-\sigma ^{0}\hat{p}_{0}+\sigma ^{i}\hat{p}_{i})\psi _{R}-m_{0}\psi _{L}=0, 
\tag{43}
\end{equation}%
and%
\begin{equation}
(\sigma ^{0}\hat{p}_{0}+\sigma ^{j}\hat{p}_{j})\psi _{L}+m_{0}\psi _{R}=0. 
\tag{44}
\end{equation}%
Thus, comparing (43) and (44) with (24) and (25) one observes that the
algebra of $\gamma ^{a}$ and $\gamma ^{\mu }$ given in (2), (18) and (19) is
the analogue of the algebra satisfied by $\sigma ^{0}$ and $\sigma ^{i}$,
namely

\begin{equation}
\sigma ^{0}\sigma ^{0}+\sigma ^{0}\sigma ^{0}=2\delta ^{00},  \tag{45}
\end{equation}

\begin{equation}
\sigma ^{0}\sigma ^{j}-\sigma ^{j}\sigma ^{0}=0,  \tag{46}
\end{equation}%
and

\begin{equation}
\sigma ^{i}\sigma ^{j}+\sigma ^{j}\sigma ^{i}=2\delta ^{ij}.  \tag{47}
\end{equation}%
This means the $\gamma ^{a}$ plays the role of $\sigma ^{0}$, while $\gamma
^{\mu }$ plays the role of $\sigma ^{i}$. This analysis may motive to look
for a $t$-time signature, rather than $s$-space signature. In fact, in the
case of $t$-time signature the formula (18) must be changed by

\begin{equation}
\gamma ^{a}\gamma ^{b}+\gamma ^{b}\gamma ^{a}=-2\delta ^{ab},  \tag{48}
\end{equation}%
At the end, starting with a Dirac equation in spacetime of $(t+s)$-signature
and imposing the cylindrical condition in the extra dimensions one must
arrive to the Elko equation in $(1+3)$-dimensions. In particular it may be
interesting to find the analogue of Elko theory in $(2+2)$-dimensions (see
Refs. [28]-[29])

From the point of view of Kaluza-Klein theory the conditions (26) and (27)
correspond to the zero mode of a compactified space. It will be interesting
for further research to consider non zero modes of the internal space. In
this direction the reference [15] dealing with brane-worlds in $5$%
-dimensions may be particularly useful.

It still remains to analyze a kind of Majorana condition for the physical
states $\psi _{a}$, namely%
\begin{equation}
C\psi _{a}=e^{i\theta }\psi _{a},  \tag{49}
\end{equation}%
where $C$ denotes a charge conjugation operator and $e^{i\theta }$ is a
phase factor. If one chooses $C\psi _{a}=\psi _{a}$ one obtains the the
self-conjugate Elko spinor or Majorana spinor, while if one requires the
condition $C\psi _{a}=-\psi _{a}$ one gets the anti-self-conjugate Elko
spinor, which is different than the convetional Majorana's choice. But this
will require to consider a Clifford algebra in a $(t+s)$-signature in a
similar way as one analyzes the Majorana-Weyl spinors in higher dimensions
[30].

In $5$-dimensions and in the Weyl representation the expression (49) implies

\begin{equation}
-\sigma ^{2}\psi _{L}^{\ast }=\psi _{R},  \tag{50}
\end{equation}%
or

\begin{equation}
\sigma ^{2}\psi _{R}^{\ast }=\psi _{L}.  \tag{51}
\end{equation}%
Consequently, up to a face, the Majorana spinor obtained from (50) and (51),
in the Weyl representation, looks like

\begin{equation}
\psi \longrightarrow \lambda =\left( 
\begin{array}{c}
-\sigma ^{2}\psi _{L}^{\ast } \\ 
\psi _{L}%
\end{array}%
\right)  \tag{52}
\end{equation}%
or

\begin{equation}
\psi \longrightarrow \rho =\left( 
\begin{array}{c}
\psi _{R} \\ 
\sigma ^{2}\psi _{R}^{\ast }%
\end{array}%
\right) .  \tag{53}
\end{equation}

It turns out that in the original construction of Elko theory [2]-[3] the
constraint (49) is considered as the starting point. For this reason when
the field equation (1) is used one finds that $\psi _{a}$ contains only $4$
complex components. In our case, in writing (15), one sees that originally $%
\psi _{a}$ has $8$ complex components, but the two approaches must be
equivalent after imposing the Majorana condition (49). Roughly speaking, in
Refs. [2]-[3] it is followed the route of (49) first and then (1), while in
our case the route is (1) (or (15)) first and then (49), but in both cases
the number of components associated with $\psi _{a}$, satisfying (1) (or
(15)) and (49), is $4$ complex components. In higher dimensions this
analysis is more complicated since besides of imposing (49) one must
consider the equations (26) and (27).

It is also worth mentioning some dimensional analysis towards a
renormalization of the theory. In general the action%
\begin{equation}
S_{(4)}=\int \mathcal{L}_{(4)}d^{4}x,  \tag{54}
\end{equation}%
must be dimensionless. For this reason, since the dimensionality of $d^{4}x$
is $-4$, one sees that each term in $\mathcal{L}_{(4)}$ must carry
dimensionality $+4$. In this context, one finds that the kinetic term of $%
\mathcal{L}_{(4)}$ implying the Dirac equation establishes that the spinor
field $\psi _{a}$ has mass dimension $3/2$. Thinking about the Lorentz
transformation of type $(A,B)$ this result is obtained considering that the
mass dimension of the spinor field is given by $1+A+B$. In fact, for the
field of the types $(1/2,0)$ and $(0,1/2)$, the expected mass dimensionality
is $3/2$. In Elko theory $\psi _{a}$ also transforms as $(1/2,0)\oplus
(0,1/2)$, but according to Refs. [2]-[3] due to non-locality the dimension
of $\psi _{a}$ is not $3/2$, but $1$ (see Ref. [31] for an alternative
results). It is worth mentioning that recently a Lagrangian approach for
mass dimension one fermions has been proposed [22]. In our case, the
situation seems different because our starting point is not the action (54)
but

\begin{equation}
S_{(5)}=\int \mathcal{L}_{(5)}d^{5}x.  \tag{55}
\end{equation}%
In this case, in order to have a dimensionless action, $\mathcal{L}_{(5)}$
must contain mass dimension $+5$, and therefore the kinetic term in $%
\mathcal{L}_{(5)}$, implying the Dirac equation in $5$ dimensions, leads to
the result that the spinor field $\psi _{a}$ must carry mass dimension $2$,
instead of $3/2$ as it is the case in $4$ dimensions. We believe that this
is an intriguing result that deserves a further research.

Finally, suppose one fix the Clifford algebra (2). One may be interested in
exploring the consequences in the Elko theory under the signature change:

\begin{equation}
(-1,1,1,1)\leftrightarrow (1,-1,-1,-1)  \tag{56}
\end{equation}%
As it has been emphasized in the the Ref. [32] only if the mass is equal to
zero the Dirac equation is invariant under (56). This conclusion may be
different in Elko theory because a massless fermion in 5-dimensions may be
massive in 4-dimensions. This seems to be an interesting question [33] which
may be a subject for further research.

\bigskip

\begin{center}
\textbf{Acknowledgments}

{\small \ }
\end{center}

I would like to thank professor D. V. Ahluwalia, as well as the two
referees, for helpful comments. This work was partially supported by
PROFAPI-UAS/2013.

\end{document}